\RequirePackage{lineno}

\documentclass[aps,prc,twocolumn,groupedaddress,showpacs,amsmath,amssymb,floatfix,superscriptaddress]{revtex4}
\usepackage{multirow}
\usepackage{graphicx}
\usepackage{times}
\usepackage{enumerate}
\usepackage{color}

\begin{document}

\title{Search for the proton decay mode $p \rightarrow \overline{\nu} K^{+}$ with KamLAND}

%

\newcommand{\tohoku}{\affiliation
	{Research Center for Neutrino Science, Tohoku University, 
	Sendai, Miyagi 980-8578, Japan}}
\newcommand{\osaka}{\affiliation
	{Graduate School of Science, Osaka University, 
	Toyonaka, Osaka 560-0043, Japan}}
\newcommand{\tokushima}{\affiliation{Faculty of Integrated Arts and Science, University of Tokushima, 
	Tokushima, 770-8502, Japan}}
\newcommand{\alabama}{\affiliation
	{Department of Physics and Astronomy, University of Alabama, 
	Tuscaloosa, Alabama 35487, USA}}
\newcommand{\lbl}{\affiliation
	{Physics Department, University of California, Berkeley, Berkeley, California 94720, USA \\ 
	and Lawrence Berkeley National Laboratory, 
	Berkeley, California 94720, USA}}
\newcommand{\hawaii}{\affiliation
	{Department of Physics and Astronomy, University of Hawaii at Manoa, 
	Honolulu, Hawaii 96822, USA}}
\newcommand{\tennessee}{\affiliation
	{Department of Physics and Astronomy, University of Tennessee, 
	Knoxville, Tennessee 37996, USA}}
\newcommand{\tunl}{\affiliation
	{Triangle Universities Nuclear Laboratory, Durham, North Carolina 27708, USA; \\
	Physics Departments at Duke University, Durham, North Carolina 27705, USA; \\
	North Carolina Central University, Durham, North Carolina 27701, USA \\
	and The University of North Carolina at Chapel Hill, Chapel Hill, North Carolina 27599, USA}}
\newcommand{\ipmu}{\affiliation
	{Kavli Institute for the Physics and Mathematics of the Universe (WPI), The University of Tokyo Institutes for Advanced Study, The University of Tokyo, Kashiwa, Chiba 277-8583, Japan}}
\newcommand{\nikhef}{\affiliation
	{Nikhef and the University of Amsterdam, Science Park, 
	Amsterdam, the Netherlands}}
\newcommand{\washington}{\affiliation
	{Center for Experimental Nuclear Physics and Astrophysics, University of Washington, 
	Seattle, Washington 98195, USA}}
\newcommand{\lsu}{\affiliation{Department of Physics and Astronomy,
	Louisiana State University, Baton Rouge, Louisiana 70803, USA}}
\newcommand{\mitc}{\affiliation{Massachusetts Institute of Technology, 
	Cambridge, Massachusetts 02139, USA}}

\newcommand{\atosakanow}{\altaffiliation
	{Present address: Department of Physics, Osaka University, 
	Toyonaka, Osaka 560-0043, Japan}}
\newcommand{\atdavisnow}{\altaffiliation
	{Present address: Physics Department, University of California, Davis, 
	Davis, California 95616, USA}}
\newcommand{\atlivermorenow}{\altaffiliation
	{Present address: Lawrence Livermore National Laboratory,
	Livermore, California 94550, USA}}
\newcommand{\deceased}{\altaffiliation {Deceased.}}

%
%
\author{K.~Asakura}\tohoku
\author{A.~Gando}\tohoku
\author{Y.~Gando}\tohoku
\author{T.~Hachiya}\tohoku
\author{S.~Hayashida}\tohoku
\author{H.~Ikeda}\tohoku
\author{K.~Inoue}\tohoku\ipmu
\author{K.~Ishidoshiro}\tohoku
\author{T.~Ishikawa}\tohoku
\author{S.~Ishio}\tohoku
\author{M.~Koga}\tohoku\ipmu
\author{R.~Matsuda}\tohoku
\author{S.~Matsuda}\tohoku
\author{T.~Mitsui}\tohoku
\author{D.~Motoki}\tohoku
\author{K.~Nakamura}\tohoku\ipmu
\author{S.~Obara}\tohoku
\author{Y.~Oki}\tohoku
\author{T.~Oura}\tohoku
\author{I.~Shimizu}\tohoku
\author{Y.~Shirahata}\tohoku
\author{J.~Shirai}\tohoku
\author{A.~Suzuki}\tohoku
\author{H.~Tachibana}\tohoku
\author{K.~Tamae}\tohoku
\author{K.~Ueshima}\tohoku
\author{H.~Watanabe}\tohoku
\author{B.D.~Xu}\tohoku
\author{Y.~Yamauchi}\tohoku
\author{H.~Yoshida}\atosakanow\tohoku

\author{A.~Kozlov}\ipmu
\author{Y.~Takemoto}\ipmu

\author{S.~Yoshida}\osaka

\author{K.~Fushimi}\tokushima

\author{C.~Grant}\atdavisnow\alabama
\author{A.~Piepke}\alabama\ipmu

\author{T.I.~Banks}\lbl
\author{B.E.~Berger}\lbl\ipmu
\author{S.J.~Freedman}\deceased\lbl
\author{B.K.~Fujikawa}\lbl\ipmu
\author{T.~O'Donnell}\lbl

\author{J.G.~Learned}\hawaii
\author{J.~Maricic}\hawaii
\author{M.~Sakai}\hawaii

\author{S.~Dazeley}\atlivermorenow\lsu
\author{R.~Svoboda}\atdavisnow\lsu

\author{L.A.~Winslow}\mitc

\author{Y.~Efremenko}\tennessee\ipmu

\author{H.J.~Karwowski}\tunl
\author{D.M.~Markoff}\tunl
\author{W.~Tornow}\tunl\ipmu

\author{J.A.~Detwiler}\washington
\author{S.~Enomoto}\washington\ipmu

\author{M.P.~Decowski}\nikhef\ipmu

\collaboration{KamLAND Collaboration}\noaffiliation

%

\date{\today}

\begin{abstract}
We present a search for the proton decay mode $p \rightarrow \overline{\nu} K^{+}$ based on an exposure of 8.97~kton-years in the KamLAND experiment. The liquid scintillator detector is sensitive to successive signals from $p \rightarrow \overline{\nu} K^{+}$ with unique kinematics, which allow us to achieve a detection efficiency of 44\%, higher than previous searches in water Cherenkov detectors. We find no evidence of proton decays for this mode. The expected background, which is dominated by atmospheric neutrinos, is $0.9 \pm 0.2$ events. The nonbackground-subtracted limit on the partial proton lifetime is $\tau / B(p \rightarrow \overline{\nu} K^{+}) > 5.4 \times 10^{32}\,\rm{years}$ at 90\% C.L.
\end{abstract}


\maketitle

\section{Introduction}
\label{section:Introduction}

Extensions of the Standard Model (SM) of particle physics, such as supersymmetric grand unified theories (SUSY-GUTs)~\cite{Wess1974,Marciano1982}, predict that the three SM coupling constants are unified at a scale of $\sim$$2 \times 10^{16}\,{\rm GeV}$. Proton decay is an indispensable probe to determine the unification scale experimentally. In SUSY-GUT models, the proton decay mode $p \rightarrow \overline{\nu} K^{+}$ is expected to be the leading process through color-triplet Higgs exchange in dimension-five operator interactions~\cite{Murayama2002}. At the same time, the predicted decay rate of $p \rightarrow e^{+} \pi^{0}$ can be highly suppressed by the masses of the decay daughters, and is consistent with the current experimental lower limits on the partial lifetime~\cite{Hirata1989,McGrew1999,Nishino2009}. 

Previous searches for the $p \rightarrow \overline{\nu} K^{+}$ decay mode were performed by the Super-Kamiokande Collaboration with a fiducial mass of 22.5\,kton~\cite{Hayato1999,Kobayashi2005,Abe2014b}. They have reported the most stringent limit of \mbox{$\tau / B(p \rightarrow \overline{\nu} K^{+}) > 5.9 \times 10^{33}\,\rm{years}$} (90\% C.L.) with a 260 kton-year exposure~\cite{Abe2014b}. In water Cherenkov detectors, the direct signal from the $K^{+}$ is not visible because its momentum is below the Cherenkov threshold, and the information on the energy released by the proton decay event is lost. So the Super-Kamiokande search relied on three detection channels in which the $K^{+}$ is identified via its decay daughters: (1) $K^{+} \rightarrow \mu^{+} \nu_{\mu}$, in delayed coincidence with deexcitation $\gamma$ rays from the remnant nucleus from $p$ decay within an oxygen nucleus; (2) $K^{+} \rightarrow \mu^{+} \nu_{\mu}$, monoenergetic muon search; and (3) $K^{+} \rightarrow \pi^{+} \pi^{0}$ search. The search with method (2) is less sensitive than the other methods due to the large atmospheric neutrino background. The background-free methods (1) and (3) show low efficiencies of 9.1\% and 10.0\%, respectively, due to backgrounds around the deexcitation $\gamma$ ray energy and the limited resolution for reconstructed pion momentum. The systematic uncertainties on the efficiencies are 22\% (Method 1), mainly arising from the dependence on nuclear models to predict the probabilities of the deexcitation $\gamma$ rays of the remnant nucleus, and 9.5\% (Method 3) from the uncertainties of the $\pi^{+}$ interaction cross section in water and the event reconstruction.

Liquid scintillator (LS) detectors, on the other hand, are expected to achieve higher detection efficiency with a lower systematic uncertainty because the scintillation light from the $K^{+}$ can be readily observed, resulting in a clear delayed coincidence signature with the signals from the $K^{+}$ decay daughters. In addition, the kinetic energy distribution of the decay $K^{+}$ measured by scintillation will peak at $\sim$$105\,{\rm MeV}$, with a spread due to nuclear effects and detector resolution, and will provide strong evidence to claim the $p \rightarrow \overline{\nu} K^{+}$ detection. The potential of a large volume LS detector was discussed in Refs.~\cite{Svoboda2003,Undagoitia2005}, and the detection efficiency of the proton decay mode $p \rightarrow \overline{\nu} K^{+}$ is expected to reach 65\%~\cite{Undagoitia2005}. However, this technique has not yet been demonstrated experimentally.

This paper demonstrates the principle of the new search method by presenting the first experimental search results for the proton decay mode $p \rightarrow \overline{\nu} K^{+}$ using a large-volume LS detector, KamLAND. KamLAND is the largest scintillation detector constructed to date, with a detector mass of 1\,kton~\cite{Gando2013b}. As discussed in Ref.~\cite{Undagoitia2005}, a significant challenge is that time difference between the prompt $K^{+}$ and delayed decay daughter signals on the same time scale as the scintillation response, requiring that the two signals be reconstructed from within an overlapping detected light profile. We show that this challenge can be overcome in practice. The signal efficiency and background studies presented in this paper are also important for a realistic estimate of the search sensitivities in future large LS detectors, and will provide useful knowledge for a comparison of the sensitivity between water Cherenkov and LS detectors in the future.

\section{KamLAND Detector}
\label{section:Detector}

KamLAND is situated 1\,km (2700\,m water equivalent) underground in the Kamioka mine in Japan. The detector consists of an 18-m-diameter stainless steel spherical shell which defines the boundary between the inner and outer detectors (ID and OD respectively).  The inner surface is instrumented with 1325 fast-timing 17-inch and 554 20-inch Hamamatsu PMTs facing inwards to a 13-m-diameter volume of LS. The density of the LS, which consists of $80\%$ dodecane, $20\%$ pseudocumene (1,2,4-trimethylbenzene), and 1.36\,g/liter PPO (2,5-diphenyloxazole), is 0.780\,g/cm$^3$ at $11.5^{\circ}$C. The LS is contained in a 135-$\mu$m-thick transparent nylon/EVOH (ethylene vinyl alcohol copolymer) balloon, which is mechanically supported by Kevlar ropes. Outside the balloon is a buffer region of dodecane and isoparaffin oils that shields the target from radioactive decays emanating from the walls of the detector and the PMTs. The OD is a cylindrical region of pure water instrumented with 225 20-inch PMTs. The OD tags cosmic-ray muons entering the detector and also absorbs neutrons from the surrounding rock.  

The 1325 17-inch PMTs alone give a photocathode coverage of $22\%$. The total PMT photocathode coverage of the inner detector is $34\%$.  Only the fast timing 17-inch PMTs were used in this analysis. When a PMT channel registers a ``hit,'' the charge as a function of time is recorded by an analog transient waveform digitizer (ATWD) chip. The KamLAND ATWDs each capture a 128 sample waveform (about 200\,ns). Each channel has two sets of three ATWDs, which record waveforms at three available gain levels, providing enough dynamic range to record signals between one and thousands of photoelectrons (p.e.'s). The dual (2 $\times$ 3) ATWD system allows signals to be captured on a second set of ATWDs during the 30\,$\mu$s required to read out the recorded waveforms from the first light detected in an event.

The data reported here are based on a total live time of 9.82~years, collected between January 10, 2003 and May 1, 2014. The data set is divided into two periods: \mbox{period I} (3.79~years live time) refers to data taken up to May 2007, at which time we embarked on a LS purification campaign that continued into 2009.  \mbox{Period II} (6.02~years live time) refers to data taken during and after the LS purification campaign. We found that the scintillation and optical parameters are different between the two periods, so we analyzed the data for each period separately, and finally the results are combined to estimate the partial proton lifetime. We use a LS target mass of 913~tons for the $p \rightarrow \overline{\nu} K^{+}$ search, resulting in a total exposure of 8.97~kton-years. Since September 2011, the \mbox{KamLAND-Zen} experiment was launched to search for neutrinoless double beta decay using a newly introduced $\beta\beta$ source, 13 tons of Xe-loaded LS in a 3.08-m-diameter inner balloon at the center of the detector~\cite{Gando2013a}. The volume fraction of Xe-LS to all LS is 0.014 and the concentration of enriched xenon gas in Xe-LS is less than 3\,wt\%, so the impact of the presence of the $\beta\beta$ source on the proton decay analysis is negligible.

\section{Simulation}
\label{section:Simulation}

Simulations were performed in order to understand the event characteristics from $p \rightarrow \overline{\nu} K^{+}$ signals and backgrounds in KamLAND. As described below, we use Monte Carlo (MC) simulation tools, an event generator for high-energy event studies, and a general particle tracker applied to the KamLAND detector. In the $p \rightarrow \overline{\nu} K^{+}$ search, the remaining background after the cosmic-ray muon rejection is dominated by atmospheric neutrinos. The atmospheric neutrino backgrounds need to be sufficiently removed by requiring delayed coincidence cuts based on the scintillation signals from the prompt $K^{+}$ and the delayed decay daughters, while maintaining a certain level of the detection efficiency. The optimum selection was studied by comparing the event characteristics of MC samples from $p \rightarrow \overline{\nu} K^{+}$ decays and atmospheric neutrinos. Finally, the signal efficiency and the background rates were evaluated from MC samples.

\subsection{$p \rightarrow \overline{\nu} K^{+}$ decay}
\label{subsection:Simulation_pdecay}

The signal efficiencies for $p \rightarrow \overline{\nu} K^{+}$ decay in KamLAND were estimated with MC simulations. We generated 10,000 event samples distributed in the complete LS volume ($1171 \pm 25\,{\rm m}^{3}$). For a free proton at rest, this two-body decay results in the back-to-back emission of the $\overline{\nu}$ and $K^{+}$ with monochromatic kinetic energies of 339\,MeV and 105\,MeV, respectively. However, the prompt signal from bound proton decay in $^{12}$C will be modified by various nuclear effects, namely the proton's Fermi motion, the nuclear binding energy, and intranuclear cascades triggered by $K^{+}$-nucleon interactions. In order to model these effects, we used a nucleon decay event generator in the \texttt{GENIE}-based MC code (version 2.8.0)~\cite{Andreopoulos2010}. \texttt{GENIE} is usually used for neutrino interactions with nuclear targets. However, the code provides models of nucleon states within a nucleus as well as intranuclear hadron transport after neutrino interactions, which are essential also for nucleon decay simulations. Previous MC studies in Ref.~\cite{Undagoitia2005} were performed under the condition that the emitted $K^{+}$ from proton decay has a kinetic energy of 105\,MeV without nuclear effects.

The distribution of momenta and binding energies of the nucleons in $^{12}$C is modeled by the spectral function of Benhar~\cite{Benhar2005}, which reproduces the experimental data of electron-$^{12}$C scattering. In the original \texttt{GENIE} code, the two-body decay of a nucleon was calculated based on a simple Fermi gas model ignoring the shell structure of the nucleus and nucleon-nucleon correlations, and the modification of the decay nucleon mass by the nuclear binding energy was absent. We implemented the mass correction $m'_{N} = m_{N} - E_{b}$ (where $E_{b}$ is nuclear binding energy) and a corresponding Fermi momentum based on the spectral function to provide a more realistic description of nucleon states in $^{12}$C.

\begin{figure}[t]
\begin{center}
\includegraphics[width=1.0\columnwidth]{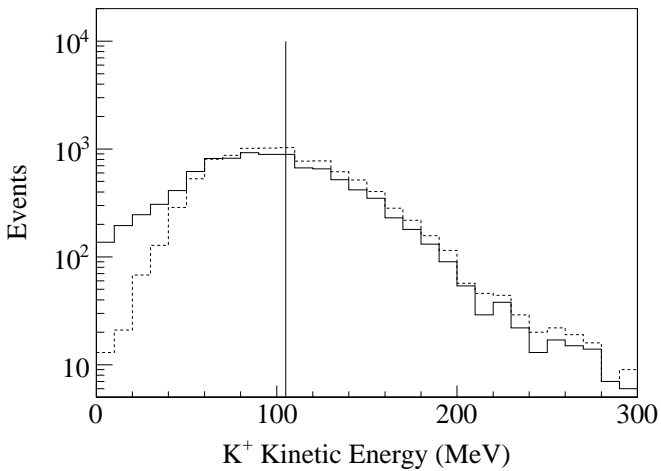}
\vspace{-0.7cm}
\end{center}
\caption[]{The $K^{+}$ kinetic energy distribution from bound proton decays in $^{12}$C with (solid line) and without (dashed line) the intranuclear cascade. The monochromatic line indicates the $K^{+}$ kinetic energy for free proton decays.
}
\label{figure:kaon_kinetic_energy}
\end{figure}

The produced hadron from nucleon decay is simulated by the intranuclear cascade model in \texttt{GENIE}. It describes the hadronic interactions with each nucleon inside the remnant nucleus before exiting. A bound proton decay of $p \rightarrow \overline{\nu} K^{+}$ in $^{12}$C produces $K^{+}$ with low momentum, below 600\,MeV/c, so only elastic scattering ($K^{+} p \rightarrow K^{+} p$ or $K^{+} n \rightarrow K^{+} n$) and charge exchange ($K^{+} n \rightarrow K^{0} p$) with nucleons in $^{11}$B can occur. In order to describe the $K^{+}$ interaction with nucleons more accurately, we modified the \texttt{GENIE} code to implement elastic scattering and charge exchange reactions calculated using the data-driven cross sections from partial-wave analysis~\cite{Hoffmann1995, Tarasov2008}. The $K^{+}$ loss probability due to charge exchange is calculated to be only 2.1\%. Figure~\ref{figure:kaon_kinetic_energy} shows the resulting $K^{+}$ kinetic energy distribution for bound proton decays in $^{12}$C. The smearing of the peak is mainly due to the Fermi momentum. For the case of $K^{+}$-nucleon interactions, we calculated the energy loss of $K^{+}$ and the energy transfer to secondaries, such as protons or neutrons, which have a different scintillation response.

\begin{figure}[b]
\begin{center}
\includegraphics[width=1.0\columnwidth]{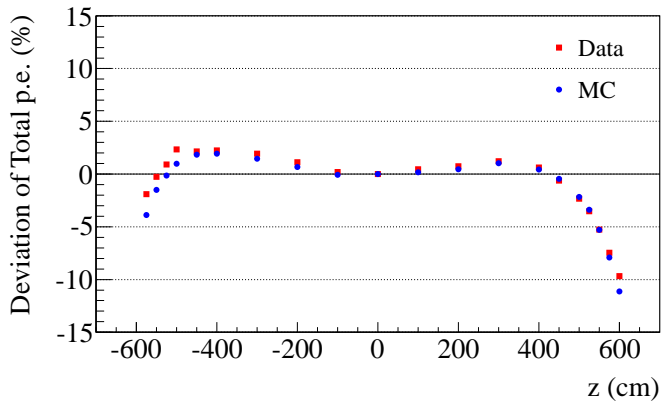}
\vspace{-0.7cm}
\end{center}
\caption[]{Fractional difference of the average total detected p.e. of $^{60}$Co events relative to the detector center for data (squares) and MC (circles). The decrease of charge around the $\pm$6\,m region is due to the absence of PMTs in the detector entrance at the top and bottom region, and shadowing by suspension ropes.}
\label{figure:charge}
\end{figure}

The initial $K^{+}$ generation with MC is followed by a set of processes for particle tracking, scintillation photon emission, and photon propagation using a separate MC code based on \texttt{Geant4}~\cite{Allison2006}. The detector response MC code was tuned to reproduce the data of radioactive source $\gamma$ rays for low energies and atmospheric neutrinos for high energies. The scintillation and optical parameters for the MC, including Birk's constant, the light yield, the attenuation length, and reemission and scattering probabilities, are determined separately for \mbox{period I} and \mbox{period II}. After the MC tuning, we found the fractional difference of the average detected number of p.e. between the measured data and the MC for central-axis radioactive source calibrations with $^{60}$Co to be less than 2\%, as shown in Fig.~\ref{figure:charge}. The time jitter and waveform for each gain channel for the 17-inch PMTs are measured with dye-laser calibration data. These are used to simulate the photon arrival time in the detector response MC output.

\begin{figure}[b]
\begin{center}
\includegraphics[width=1.0\columnwidth]{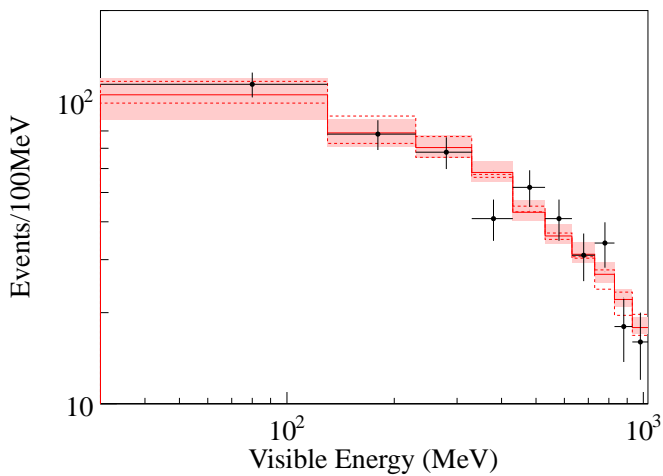}
\vspace{-0.7cm}
\end{center}
\caption[]{Reconstructed energy spectrum of atmospheric neutrino candidates (black dots) within a 5-m-radius volume. The atmospheric neutrino MC (solid line) is normalized to the live time of data. The neutrino interaction uncertainties evaluated by the \texttt{GENIE} code are indicated by the shaded region. The expectations for the linear energy scaling with $\pm 1\sigma$ are also shown (dashed lines).}
\label{figure:energy_candidates}
\end{figure}

\begin{figure*}[t]
\includegraphics[width=0.65\columnwidth]{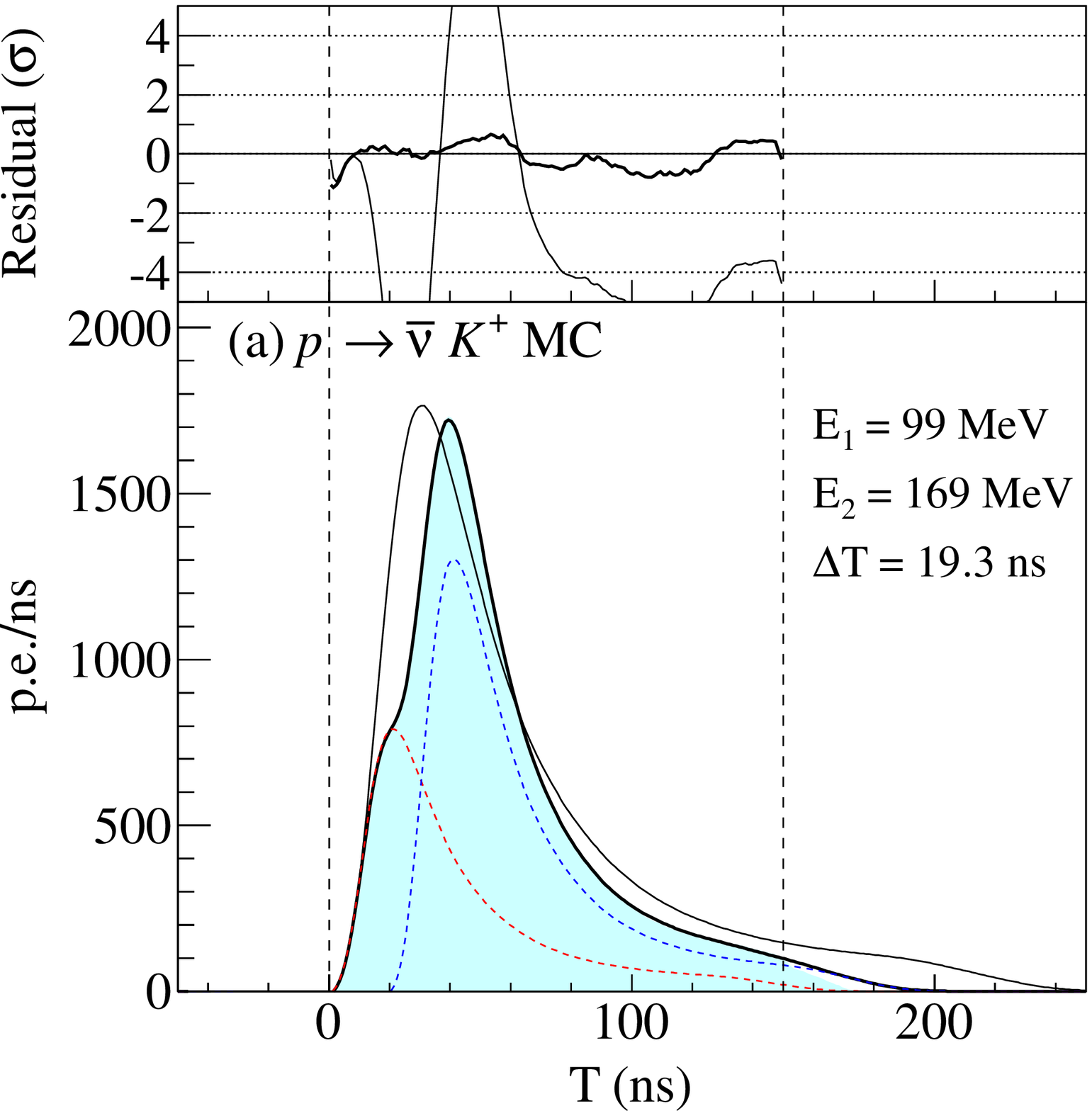}
\hspace{0.3cm}
\includegraphics[width=0.65\columnwidth]{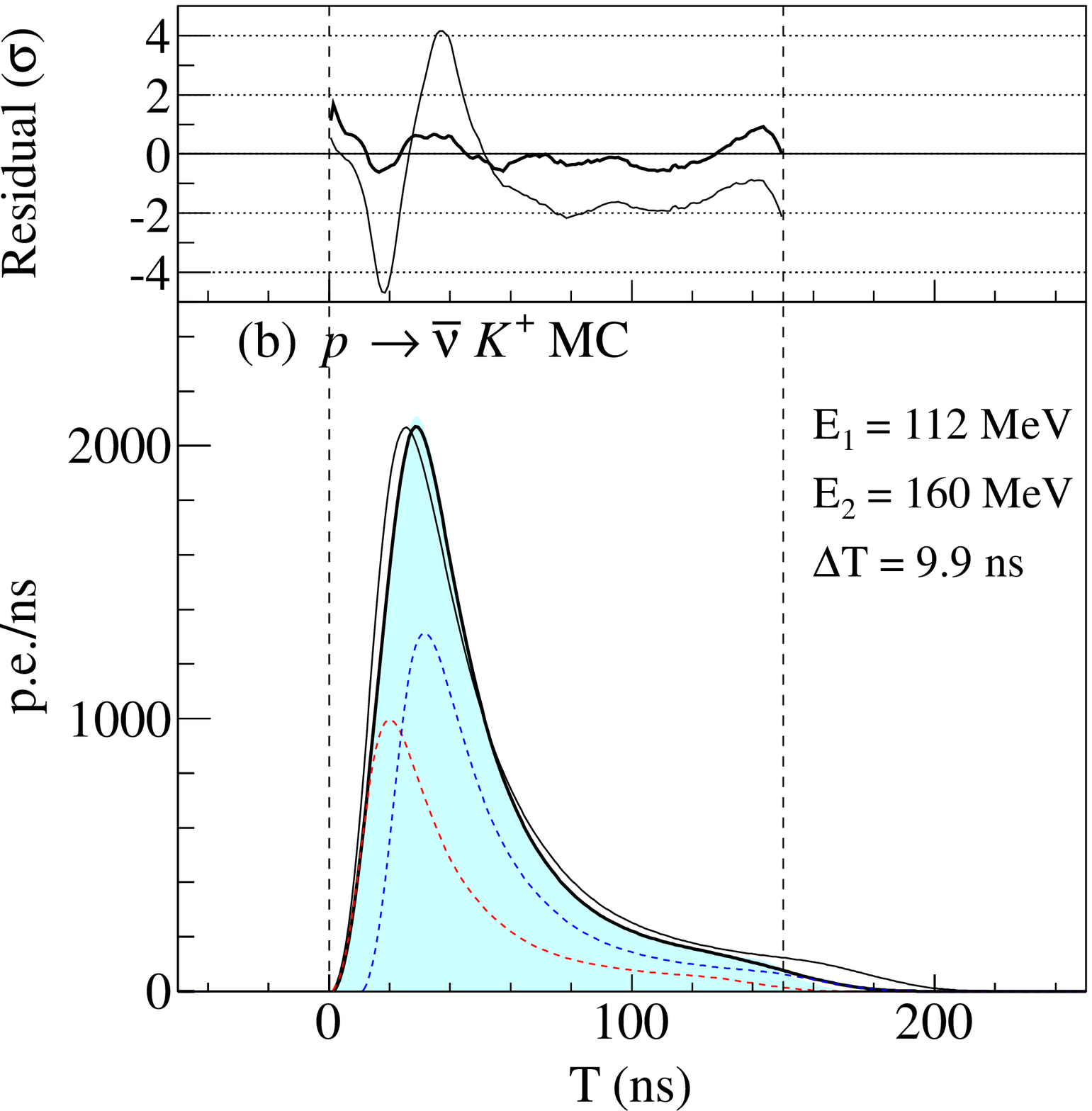}
\hspace{0.3cm}
\includegraphics[width=0.65\columnwidth]{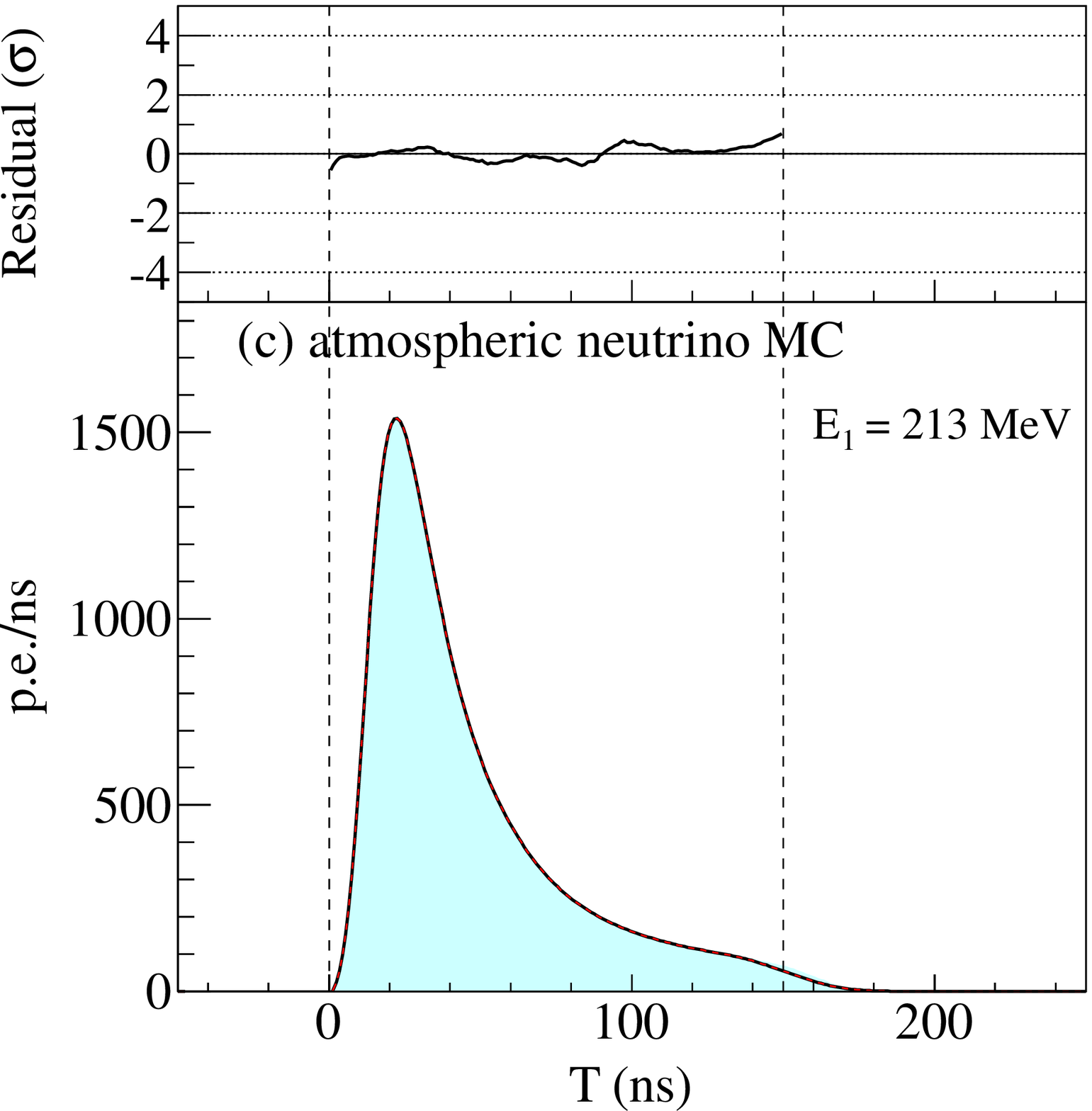}
\caption[]{Typical waveform shape (shaded) expected from (a) and (b) $p \rightarrow \overline{\nu} K^{+}$ MC example events and (c) an atmospheric neutrino MC event, together with the best fit curves from the multipulse fit (solid thick line) and from the single-pulse fit (solid thin line). In the upper panel, the residuals from the best fit are shown to identify multipulse events. The fitted energies and, for the multipulse fits, time differences between the first pulse from the $K^{+}$ (red dashed line) and the second pulse from the $\mu^{+}$ decay daughter of the $K^{+}$ (blue dashed line) are also shown. The shape parameters (see text) for each fit are (a) $\log(L_{shape}) = 0.06$ and (b) $\log(L_{shape}) = 0.02$.}
\label{figure:multi_pulse_fit_MC}
\end{figure*}

\begin{figure*}[t]
\includegraphics[width=0.65\columnwidth]{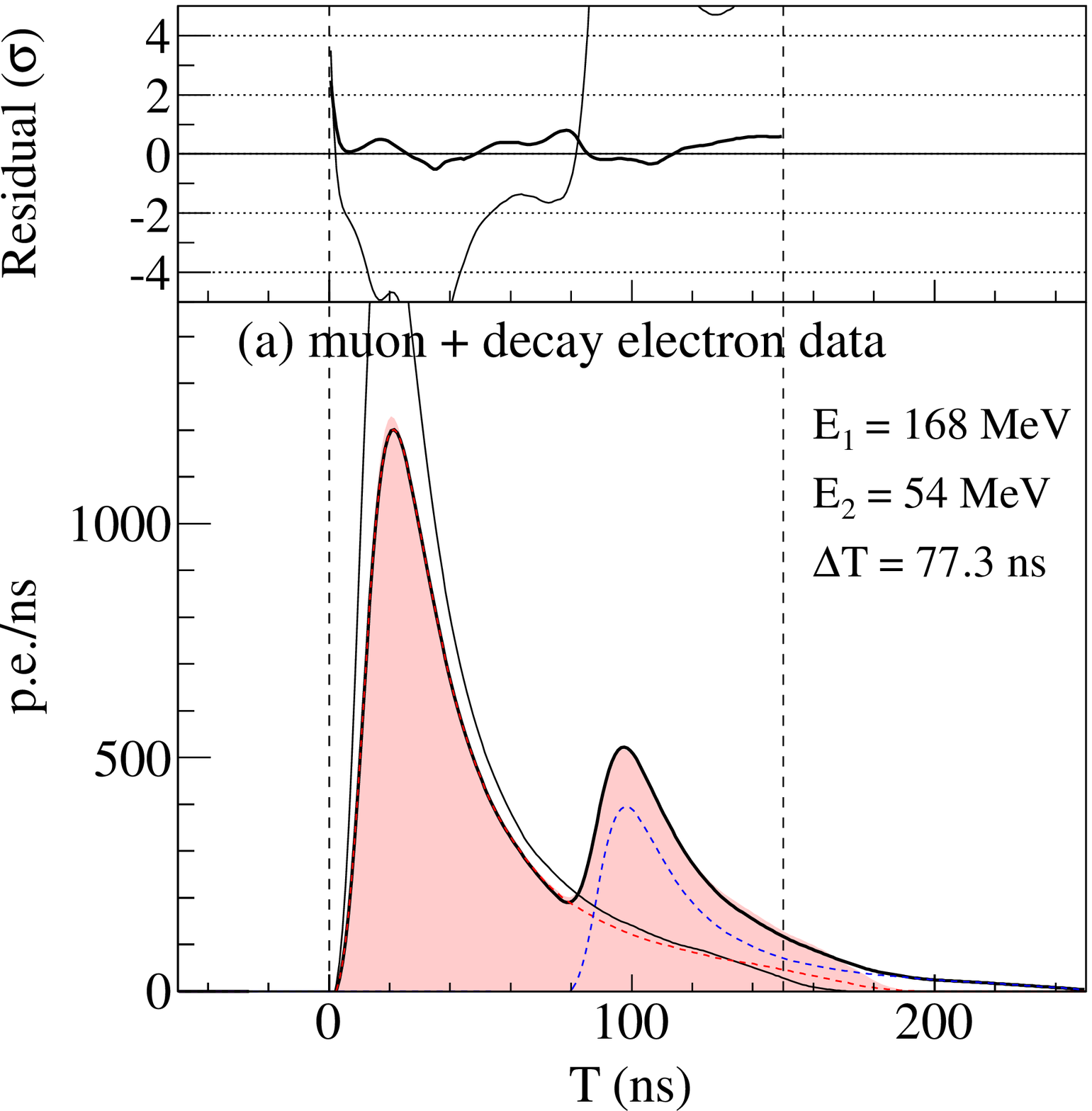}
\hspace{0.3cm}
\includegraphics[width=0.65\columnwidth]{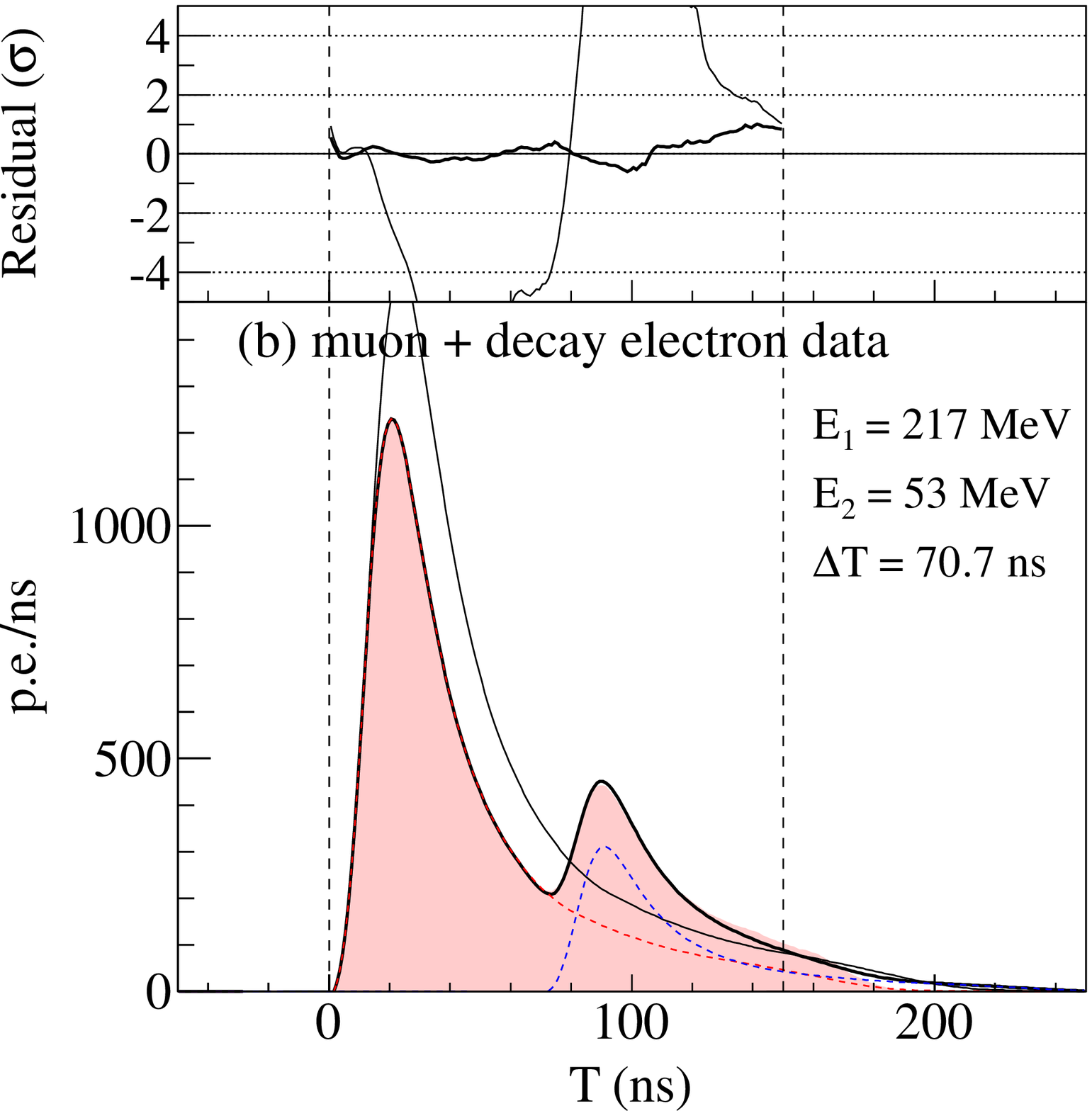}
\hspace{0.3cm}
\includegraphics[width=0.65\columnwidth]{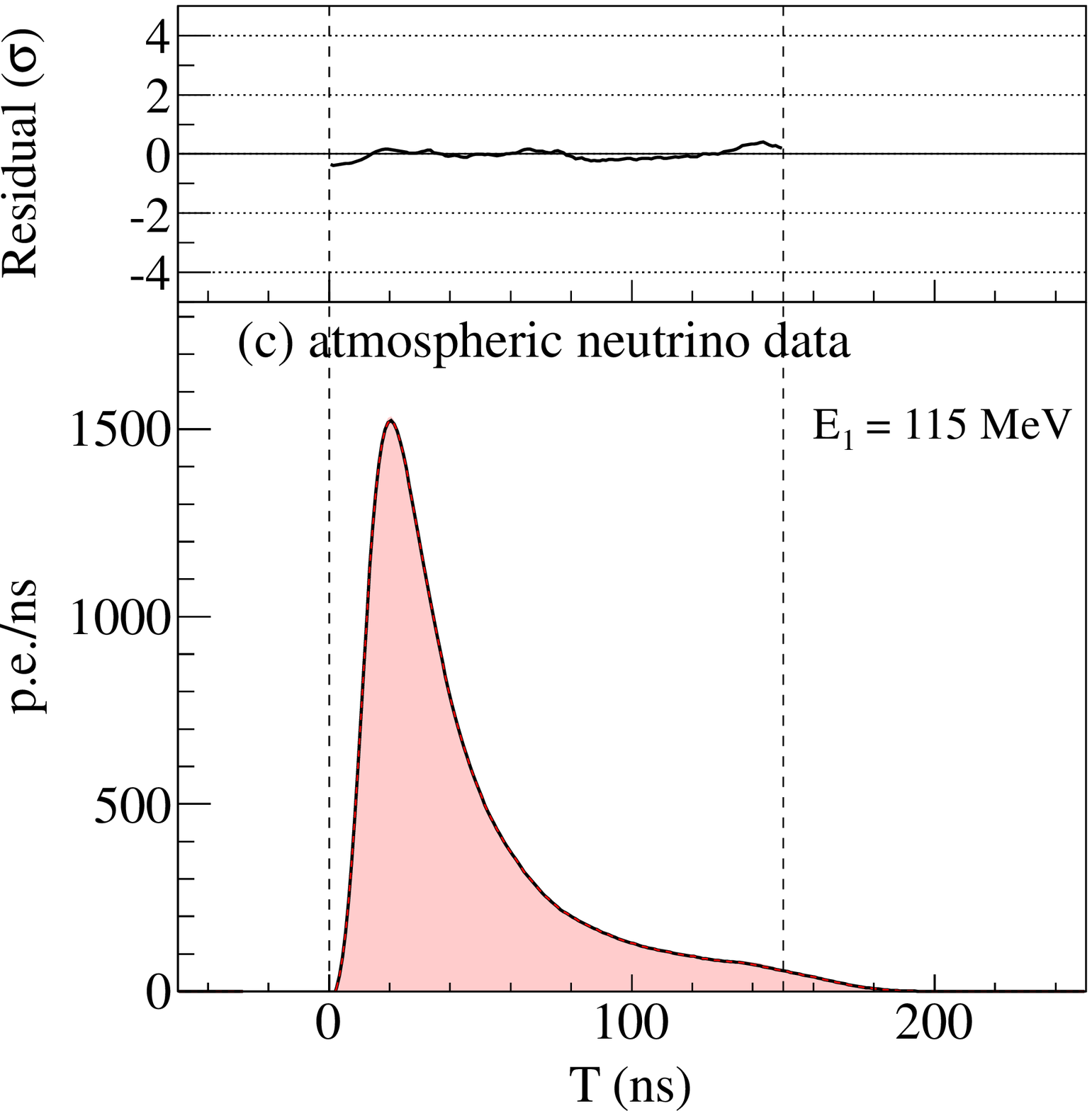}
\caption[]{Typical waveform shape (shaded) from: (a) and (b) muon $+$ decay electron data and (c) atmospheric neutrino data, together with the best fit curves from the multipulse fit (solid thick line) and from the single-pulse fit (solid thin line). In the upper panel, the residuals from the best fit are shown to identify multipulse events. The fitted energies and, for the multipulse fits, time differences between the first pulse from muons (red dashed line) and the second pulse from decay electrons (blue dashed line) are shown.}
\label{figure:multi_pulse_fit_data}
\end{figure*}

\subsection{Atmospheric neutrino background}
\label{subsection:Simulation_atmospheric_neutrino}

The backgrounds from atmospheric neutrinos are estimated with MC simulations. We used the \texttt{GENIE}-based MC event generator assuming the atmospheric neutrino flux of Ref.~\cite{Honda2011}. Three-flavor neutrino oscillations based on the best fit parameters given in Ref.~\cite{Olive2014} were taken into account. The detector simulation after the event generation is performed by the \texttt{Geant4}-based MC code described above. We produced 20,000 atmospheric neutrino MC events in KamLAND which corresponds to about 100 years live time.

The reactions are dominated by quasielastic scattering below 1 GeV, producing a charged lepton in a charged-current interaction or a recoil nucleon in neutral current scattering. While those reactions overlap with the $p \rightarrow \overline{\nu} K^{+}$ signal in visible energy, the background can be effectively suppressed by the delayed coincidence requirement. The background rate from the misidentification of single-pulse atmospheric neutrino hit patterns as multipulse events is estimated in the following section. 

Other potential background sources are kaons produced by the neutrino interactions. The \texttt{GENIE} code simulates kaon production via resonances based on the Rein-Sehgal model~\cite{Rein1981}. These reactions conserve strangeness, resulting in pair production of kaons and hyperons in the final state, such as $\nu_{\mu} + n \rightarrow \mu^{-} + \Lambda + K^{+}$. Based on the \texttt{GENIE} calculation, the number of events below 2\,GeV with a $K^{+}$ produced via a resonance is expected to be only 0.2 in the data set. On the other hand, nonresonant kaon production, such as $\nu_{\mu} + p \rightarrow \mu^{-} + p + K^{+}$, is not considered in the \texttt{GENIE} code. Although experimental confirmation of such reactions is absent, these cross sections are calculable theoretically~\cite{Alam2010}. While they are comparable with the cross sections for resonant $K^{+}$ production, due to the lower reaction thresholds, nonresonant kaon production could be a more important background in the $p \rightarrow \overline{\nu} K^{+}$ search. However, based on the cross sections in Ref.~\cite{Alam2010}, the number of events below 2\,GeV is expected to be only 0.14 in the data set. Therefore, the background contributions from both resonant and nonresonant kaon production are not significant for the current $p \rightarrow \overline{\nu} K^{+}$ search in KamLAND.

\section{Analysis}
\label{section:Analysis}

Clear identification of the two scintillation signals from the prompt $K^{+}$ and the delayed decay daughters requires the multipulse fit to the PMT hit-time pattern as discussed later in this section. However, such fits to all the recorded events take a lot of computing time. Therefore, we apply three selection criteria prior to the delayed coincidence fitting: (i) the reconstructed energy ($E$) based on the PMT charge recorded within a restricted 150\,ns event window is required to be more than 30\,MeV; (ii) the number of OD PMT hits is less than 5; and (iii) the small contribution of events in time coincidence with the neutrino beam of the T2K experiment~\cite{Abe2014a} are rejected. Unlike the low-energy radioactive decays that comprise the bulk of the KamLAND data, most of the $p \rightarrow \overline{\nu} K^{+}$ decays are expected to deposit a much higher energy, typically peaked at 260\,MeV including short-lived decay daughter contributions. So the radioactivity background can be effectively suppressed with (i). The event rate of cosmic-ray muons which pass through the ID is about 0.3\,Hz, but those muons can be rejected by the OD tagging of (ii). The inefficiency of the OD is estimated to be about 0.2\%, and the remaining muons are rejected by the PMT hit-time distribution, discussed later. The events remaining after application of these selection criteria are dominated by atmospheric neutrinos. Figure~\ref{figure:energy_candidates} shows the reconstructed energy spectrum of candidate events within a 5-m-radius volume after the prior selections. The observed spectrum is consistent with the expectation from the MC simulation. The uncertainty of the energy scale model, discussed in Sec.~\ref{section:Efficiency}, is constrained from these data.

These high-energy candidates were used for the $p \rightarrow \overline{\nu} K^{+}$ signal search. In order to identify the successive signals from $K^{+}$ and its decay daughters over a short time, we decompose one event into multiple pulses based on a binned-$\chi^{2}$ fit to the sum of the waveforms from all hit 17-inch PMTs. The waveform is a function of the photon arrival time after subtracting the time of flight (TOF) from the vertex to each PMT. The average waveform for high-energy candidates is used as the reference pulse for the fit. Considering the energy and position dependence of the waveform shape, the reference pulse was prepared for each energy-position bin. The binned-$\chi^{2}$ for $M$ pulses is defined as
\begin{eqnarray}
\label{equation:chi2}
\chi^{2}_{M} = \sum_{i} \left[ f (T_{i}) - \sum_{j}^{M} \frac{E_{j}}{E} \phi(T_{i} - \Delta T_{j}) \right]^{2} / \sigma^{2} (T_{i}), 
\end{eqnarray}
where $f(T_{i})$ is the observed waveform sum for the $i$th bin (bin width is 1\,ns) as a function of $T$ ($0 < T < 150\,{\rm ns}$), $\phi (T_{i})$ is the reference pulse, $E$ is the total visible energy estimated from the observed charge within the 150\,ns event window, $E_{j}$ is the energy contribution for the $j$th pulse, and $\Delta T_{j}$ is the time difference between the first and the $j$th pulse ($\Delta T_{1} \equiv 0$). $\sigma^{2} (T_{i})$ is calculated by
\begin{eqnarray}
\label{equation:sigma2}
\sigma^{2} (T_{i}) = \sum_{j}^{M} \left[ \frac{E_{j}}{E} \Delta \phi(T_{i} - \Delta T_{j}) \right]^{2},
\end{eqnarray}
where $\Delta \phi (T_{i})$ is the statistical fluctuation (1$\sigma$ deviation) of the reference pulse evaluated by the MC simulation. Since atmospheric neutrino interactions can produce electromagnetic showers, the waveform shape may change. In order to account for such fluctuations, the parameter $\alpha$ is introduced to scale the time ($T \rightarrow \alpha T$) of the first pulse and is floated to fit the peak time within the uncertainty. In the main branching ratio of the $K^{+}$ decay channel, $K^{+} \rightarrow \mu^{+} \nu_{\mu}$ (63.55\%), the event can contain up to three pulses, where the second pulse accounts for the energy deposition of the $\mu^{+}$, and the third pulse is for the $e^{+}$ decay daughter of the $\mu^{+}$. However, due to the short digitization window (a few hundred ns) relative to the muon lifetime (2.2 $\mu$s), and due to ATWD dead time and misfires resulting from baseline instability in very high PMT charge events, the separated $e^{+}$ event is difficult to identify. So in this analysis the $e^{+}$ tagging is not used and we restrict the multipulse fit to $M$ = 1 or 2 only. In the second-largest branching ratio, $K^{+} \rightarrow \pi^{+} \pi^{0}$ (20.66\%), the second pulse is for $\pi^{+} \pi^{0}$ decay daughters of the $K^{+}$ with the instantaneous electromagnetic $\pi^{0}$ decay, resulting in a higher energy deposit than the main branch. The subsequent $\mu^{+}$ decay daughter of the $\pi^{+}$ has a negligible impact on the fit because of the small energy deposit of 4.2\,MeV. Similarly, we use the multipulse fit to search for all the $K^{+}$ decay channels. The strongest decay branches and the corresponding kinetic energy sums that comprise the delayed scintillation signal are listed in Table~\ref{table:branch}.

\begin{table}[t]
\begin{center}
\caption{\label{table:branch}Main decay modes of $K^{+}$ in descending order of branching ratio~\cite{Olive2014}. The prompt $K^{+}$ signal is followed by delayed signals from decay daughters of the $K^{+}$. Their kinetic energies which contribute to instantaneous scintillation are summed up.}
\begin{tabular}{@{}*{3}{lcc}}
\hline
\hline
Decay mode & ~~Branching ratio (\%)~~ & Kinetic energy sum (MeV) \\
\hline
$K^{+} \rightarrow \mu^{+} \nu_{\mu}$ & $63.55 \pm 0.11$ & 152 \\
$K^{+} \rightarrow \pi^{+} \pi^{0}$ & $20.66 \pm 0.08$ & 354 \\
$K^{+} \rightarrow \pi^{+} \pi^{+} \pi^{-}$ & $5.59 \pm 0.04$ & 75 \\
$K^{+} \rightarrow \pi^{0} e^{+} \nu_{e}$ & $5.07 \pm 0.04$ & 265$-$493 \\
$K^{+} \rightarrow \pi^{0} \mu^{+} \nu_{\mu}$ & $3.353 \pm 0.034$ & 200$-$388 \\
$K^{+} \rightarrow \pi^{+} \pi^{0} \pi^{0}$ & $1.761 \pm 0.022$ & 354 \\
\hline
\hline
\end{tabular}
\vspace{-0.5cm}
\end{center}
\end{table}

Figures~\ref{figure:multi_pulse_fit_MC}(a) and~\ref{figure:multi_pulse_fit_MC}(b) show multipulse fit results for typical $p \rightarrow \overline{\nu} K^{+}$ MC events. The plots are labeled with the best fit energies of the prompt $K^{+}$ and delayed $\mu^{+}$ scintillation pulses ($E_{1,2}$), and the time difference between them ($\Delta T$). To demonstrate the performance of the fit, Figs.~\ref{figure:multi_pulse_fit_data}(a) and~\ref{figure:multi_pulse_fit_data}(b) show multipulse fit results for typical multipulse events from muon decay events recorded by KamLAND, labeled with best fit values for the reconstructed muon and Michel electron energies ($E_{1}$ and $E_{2}$, respectively) and the decay time ($\Delta T$). We also show the performance of the fit on typical atmospheric neutrino events in both MC and data in Figs.~\ref{figure:multi_pulse_fit_MC}(c) and~\ref{figure:multi_pulse_fit_data}(c), which are best fit with $M=1$.

We require the following criteria to select the $p \rightarrow \overline{\nu} K^{+}$ signal:

\begin{enumerate}[(1)]
\item The distance from the center of the detector to the reconstructed vertex ($R$) is less than 6.5\,m, which corresponds to the full LS volume.
\item The PMT hit-time distribution is required to be more pointlike than tracklike to reject cosmic-ray muons which have long tracks in the ID. Specifically, the time-based \mbox{\it vertex}-$\chi^{2}$ ($\chi^{2}_{T}$), which compares the observed PMT hit-time distribution as a function of distance to the event to the expected distribution, is required to be less than 1.2 (pointlike) for the full volume, or to lie between 1.2 and 5.0 (short-tracklike) for $z < 5.0\,{\rm m}$ and $x^2+y^2 < 25\,{\rm m}^{2}$.
\item The best fit time difference between the first pulse and the second pulse ($\Delta T$) is between 7 and 100\,ns, considering the $K^{+}$ lifetime of 12.4\,ns.
\item The best fit energy of the first pulse ($E_{1}$) is between 30 and 300\,MeV, the second pulse energy ($E_{2}$) is between 70 and 600\,MeV, as constrained by the decay kinematics.
\item The waveform shape is multipulselike to reject single-pulse events. We use a waveform shape parameter defined as $\log(L_{shape}) \equiv \log(\Delta \chi^{2}) - \lambda$, where \mbox{$\Delta \chi^{2} = \chi^{2}_{1, min} - \chi^{2}_{2, min}$} (where $\chi^{2}_{M, min}$ is the best fit $\chi^{2}$ for $M$ pulses) and $\lambda$ is the mean of the $\log(\Delta \chi^{2})$ distributions evaluated with the $p \rightarrow \overline{\nu} K^{+}$ MC as a function of $\Delta T$. $\log(L_{shape})$ is required to be more than $-0.5$.
\end{enumerate}

\begin{figure}[t]
\includegraphics[width=\columnwidth]{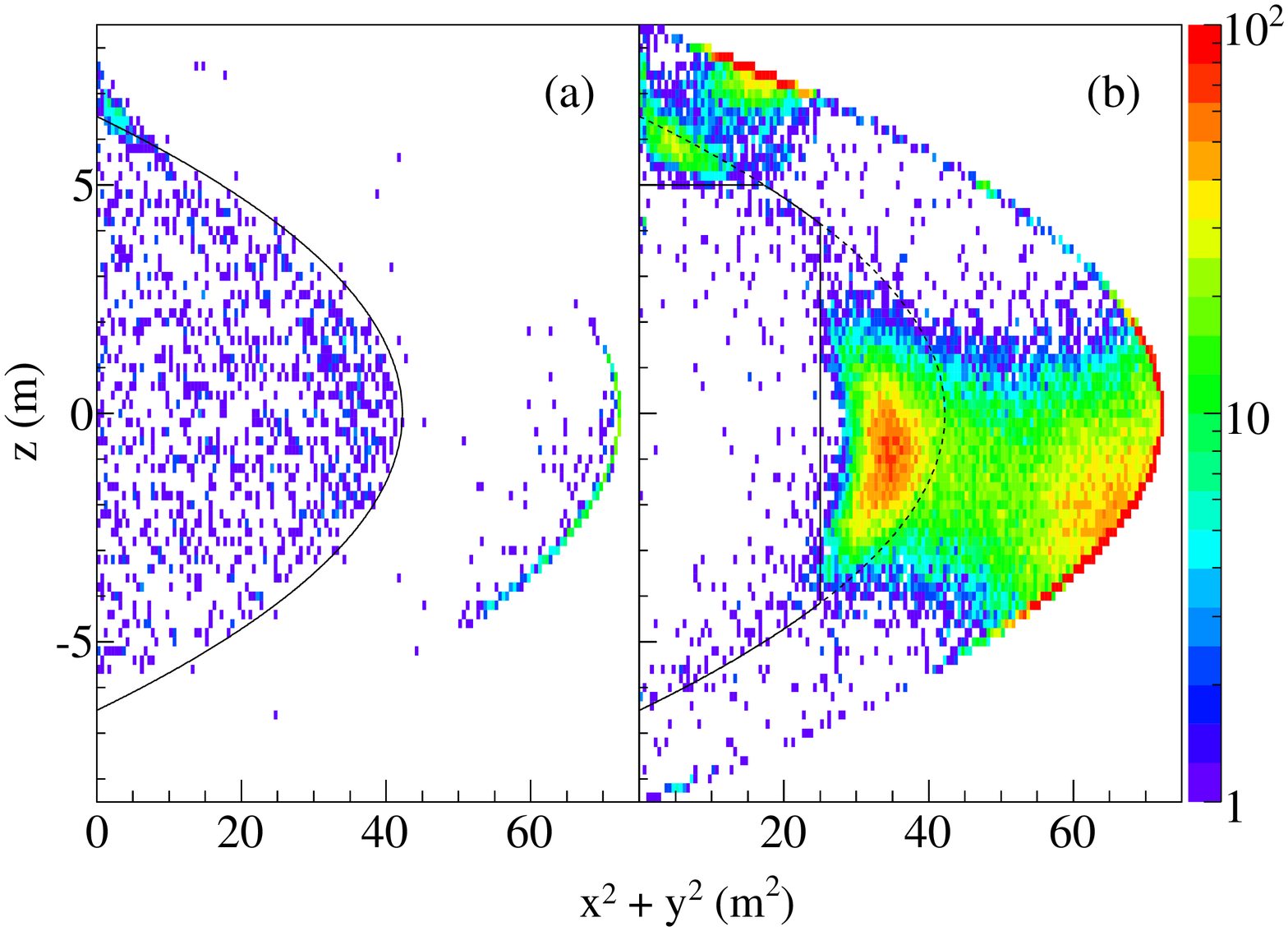}
\caption[]{Vertex distribution of atmospheric neutrino candidates in the energy range of 30$-$1000\,MeV for (a) pointlike events ($\chi^{2}_{T} < 1.2$) and (b) short-tracklike events ($1.2 \leq \chi^{2}_{T} < 5.0$).}
\label{figure:vertex}
\end{figure}

Criterion 2 is essential for removing the background from OD untagged cosmic-ray muons. These events concentrate in the top and equator regions of the detector because the water Cherenkov active volume outside these regions is relatively thin. Muons traversing these regions also tend to have short paths. Figure~\ref{figure:vertex} shows the vertex distributions of atmospheric neutrino candidates in the energy range of 30$-$1000\,MeV. The requirement of a pointlike hit timing distribution is effective at removing this OD untagged muon background within the full KamLAND volume. To maintain efficiency for $K^{+}$ decays at larger $\Delta T$ this fit must be relaxed, but this can only be done within the tighter fiducial volume defined by (2).

\begin{figure*}[t]
\includegraphics[width=1.8\columnwidth]{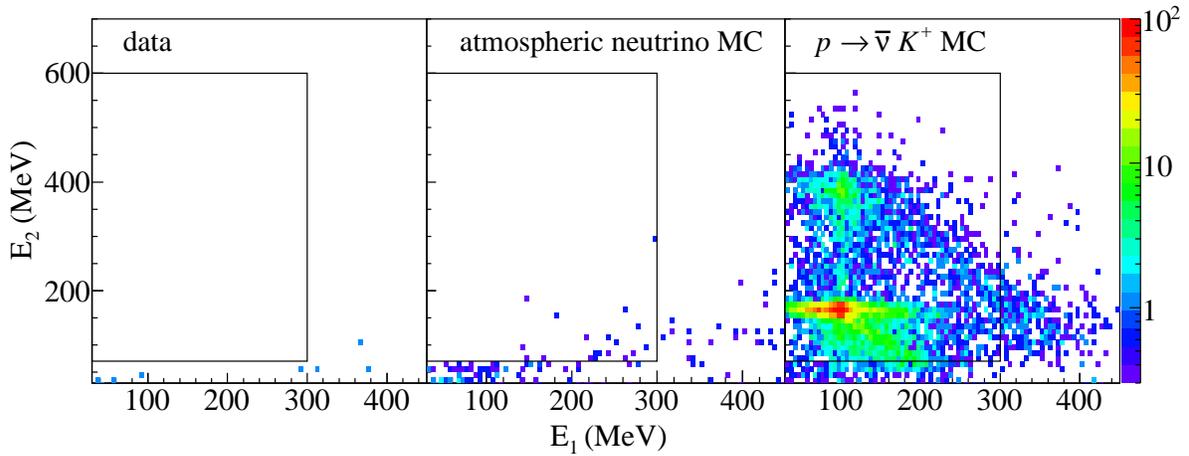}
\caption[]{Distribution of best fit energies of the two pulses $E_{1}$ and $E_{2}$ for data (left), atmospheric neutrino MC events (middle), and $p \rightarrow \overline{\nu} K^{+}$ MC events (right). $E_{1}$ provides an estimate of the $K^{+}$ kinetic energy, as illustrated in Fig.~\ref{figure:kaon_kinetic_energy}, and $E_{2}$ the kinetic energy sum for decay daughters of the $K^{+}$, summarized in Table~\ref{table:branch}. The box shows the selection cuts for $E_{1}$ and $E_{2}$. The detection efficiency and the expected number of background events were estimated from the MC event distributions for $p \rightarrow \overline{\nu} K^{+}$ (right) and atmospheric neutrinos (middle), respectively.}
\label{figure:energy}
\end{figure*}

\begin{figure}[t]
\includegraphics[width=0.75\columnwidth]{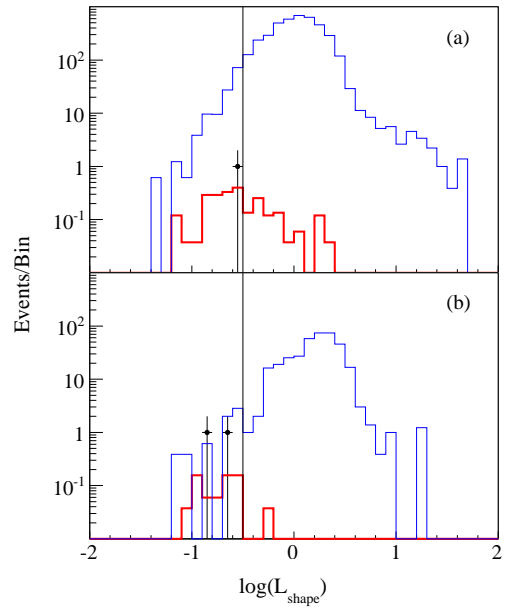}
\vspace{0.5cm}
\caption[]{Shape parameter ($L_{shape}$) distributions of data (black dot), atmospheric neutrino MC (thick red line), and $p \rightarrow \overline{\nu} K^{+}$ MC (thin blue line) for (a) pointlike events ($\chi^{2}_{T} < 1.2$) and (b) short-tracklike events ($1.2 \leq \chi^{2}_{T} < 5.0$). The contribution of the atmospheric neutrino background is normalized to the live time of data.}
\label{figure:shape_parameter}
\end{figure}

Figure~\ref{figure:energy} shows the distributions of best fit energies of the two pulses ($E_{1}$, $E_{2}$) for data, atmospheric neutrino MC events, and $p \rightarrow \overline{\nu} K^{+}$ MC events, after all cuts except (4). The main background sources are misidentified single-pulse events from atmospheric neutrinos due to the limited multipulse separation power in the fit. The multipulse fit result for $p \rightarrow \overline{\nu} K^{+}$ MC events (Fig.~\ref{figure:energy}, right panel) should reproduce the kinetic energy distribution for $K^{+}$ ($E_{1}$) illustrated in Fig.~\ref{figure:kaon_kinetic_energy}, and for decay daughters of the $K^{+}$ ($E_{2}$). In the $E_{2}$ distribution, there are two bands corresponding to stopped $K^{+}$ decays, $K^{+} \rightarrow \mu^{+} \nu_{\mu}$ with a branching ratio of 63.55\% (the kinetic energy of the $\mu^{+}$ is 152\,MeV) and  $K^{+} \rightarrow \pi^{+} \pi^{0}$ with a branching ratio of 20.66\% (the sum of kinetic energies of the $\pi^{+}$ and the $\pi^{0}$-decay-induced $\gamma$ rays is 354\,MeV). The difference between the visible energies and the real energies is explained by the nonlinear scintillation response, primarily Birk's quenching. The anticorrelation between $E_{1}$ and $E_{2}$ is due to the multipulse separation uncertainty for the case of the smaller time difference between the first and second pulses or to contributions from other $K^{+}$ decay modes. Figure~\ref{figure:shape_parameter} shows the distribution of the shape parameter for data, atmospheric neutrino MC, and $p \rightarrow \overline{\nu} K^{+}$ MC, after all cuts except (5). The atmospheric neutrino backgrounds from misidentification of single-pulse events tend to have lower $L_{shape}$ as expected. The cut values for $E_{1}$ and $L_{shape}$ were chosen using MC only maximizing the sensitivity of the search with respect to the resulting variation of the detection efficiencies, the expected number of background events, and the expected statistical uncertainty. The expected number of atmospheric neutrino background events is $0.9 \pm 0.2$ in total. No data remain after all cuts.

\begin{table}[t]
\begin{center}
\caption{\label{table:efficiency}Detection efficiencies estimated from $p \rightarrow \overline{\nu} K^{+}$ MC at each event reduction step for the five selection criteria.}
\begin{tabular}{@{}*{3}{lcc}}
\hline
\hline
\hspace{3.8cm} & ~~~~~~~Period I~~~~~~~ & ~~~~~~~Period II~~~~~~~ \\
\hline
(1) Vertex cut ($x, y, z$) & 0.979 & 0.978 \\
(2) Pointlike cut ($\chi^{2}_{T}$) & 0.935 & 0.932 \\
(3) Time cut ($\Delta T$) & 0.658 & 0.624 \\
(4) Energy cut ($E_{1}, E_{2}$) & 0.454 & 0.447 \\
(5) Shape cut ($L_{\rm shape}$) & 0.444 & 0.432 \\
\hline
\hline
\end{tabular}
\vspace{-0.5cm}
\end{center}
\end{table}

\section{Efficiencies and Systematic Uncertainties}
\label{section:Efficiency}

\begin{figure}[t]
\includegraphics[width=1.0\columnwidth]{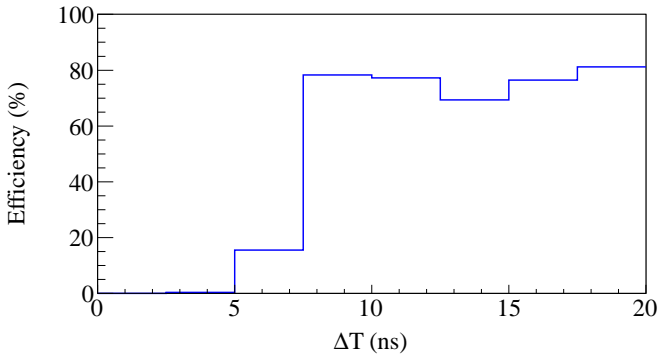}
\caption[]{Detection efficiency as a function of $\Delta T$ estimated from \mbox{$p \rightarrow \overline{\nu} K^{+}$} MC in \mbox{period I}. The decrease for $\Delta T < 7\,{\rm ns}$, due to poor resolution of the multipulse fit, is the primary source of inefficiency for the $p \rightarrow \overline{\nu} K^{+}$ search.}
\label{figure:Efficiency}
\end{figure}

Table~\ref{table:efficiency} shows the detection efficiencies estimated from \mbox{$p \rightarrow \overline{\nu} K^{+}$} MC at each event reduction step for the five selection criteria. The MC statistical uncertainty is less than 0.5\%. The largest inefficiencies, caused by the selection criteria (3) and (4), are highly dependent on the multipulse separation power in the fit. The decrease of the detection efficiencies for the case of $K^{+}$ decay within a short time is illustrated in Fig.~\ref{figure:Efficiency}. The total efficiencies for $p \rightarrow \overline{\nu} K^{+}$ in the LS are estimated to be $0.444 \pm 0.053$ for \mbox{period I} and $0.432 \pm 0.050$ for \mbox{period II}, considering the systematic uncertainties on the detection efficiency listed in Table~\ref{table:systematic}.

The number of protons is calculated based on the total LS volume ($1171 \pm 25\,{\rm m}^{3}$) measured by flow meters during detector filling, the LS density (0.780\,g/cm$^3$), and the hydrogen-to-carbon ratio (1.97) verified by elemental analysis. The nitrogen-to-carbon and oxygen-to-carbon ratios are negligibly small ($<6 \times 10^{-4}$). The 913~tons of LS mass contain $7.74 \times 10^{31}$ free protons in hydrogen and $2.33 \times 10^{32}$ bound protons in carbon, and thus $3.11 \times 10^{32}$ protons in total. The uncertainty on the number of protons is 2.1\%, determined from the LS volume measurement. 

The detection efficiency calculation from \mbox{$p \rightarrow \overline{\nu} K^{+}$} MC has associated systematic uncertainties. The MC code considers $K^{+}$-nucleon interactions in $^{12}$C, which cause the $K^{+}$ loss with a probability of 2.1\% assuming the nuclear model given in Sec.~\ref{subsection:Simulation_pdecay}. We conservatively include the size of this correction as a systematic uncertainty. The detector energy scale in the low-energy data is well calibrated by the radioactive source $\gamma$ rays, but in the energy scale of this analysis we must rely on the constraints provided by the fit to the spectral shape of atmospheric neutrino candidates up to 1\,GeV shown in Fig.~\ref{figure:energy_candidates}. In this fit we float the linear energy scale, the normalization factors, and the two systematic parameters provided in the GENIE code: the axial mass for charged-current quasielastic scattering and the axial mass for neutral-current elastic scattering, which are inputs for the calculation of the nucleon form factor in nuclei. They are constrained to lie within their uncertainties. The energy scale factor is fitted to $0.95^{+0.11}_{-0.17}$ at $\pm 1\sigma$ as indicated in Fig.~\ref{figure:energy_candidates}, which corresponds to a 3.6\%/1.9\% uncertainty in the detection efficiency for Period I/II. 

The primary source of systematic uncertainty is a possible waveform deformation due to uncertainties in the scintillation process. The scintillation time profile is determined from the ionization density of charged particles interacting in the LS; thus the waveform shape depends on the type of particle interacting and its energy. As there are no waveform calibration samples for $K^{+}$ scintillation, we computed the efficiencies from \mbox{$p \rightarrow \overline{\nu} K^{+}$} MC for two extreme cases in which the scintillation time profile is constructed from the low-energy data for a $\gamma$ ray source and for a neutron source. The ionization density of a $K^{+}$ from proton decay is expected to be in between these two cases. We found a detection efficiency difference between these two of 11.1\% and assign this as a systematic uncertainty. The total systematic uncertainty on the lifetime calculation for $p \rightarrow \overline{\nu} K^{+}$ is 12.0\%/11.6\% for Period I/II, from adding the uncertainty sources summarized in Table~\ref{table:systematic} in quadrature.

\begin{table}[t]
\begin{center}
\caption{\label{table:systematic}Estimated systematic uncertainties of the lifetime calculation for $p \rightarrow \overline{\nu} K^{+}$. The dominant uncertainty comes from the waveform modeling (11.1\%) for the $K^{+}$ signal in the detection efficiency calculation.}
\begin{tabular}{@{}*{3}{lcc}}
\hline
\hline
Source & ~~~~~~Period I~~~~~~ & ~~~~~~Period II~~~~~~ \\
\hline
Number of protons & 2.1\% & 2.1\% \\
Detection efficiency & & \\
~~~Nuclear model & 2.1\% & 2.1\% \\
~~~Energy scale & 3.6\% & 1.9\% \\
~~~Waveform model & 11.1\% & 11.1\% \\
\hline
Total systematic uncertainty~~~~~~ & 12.0\% & 11.6\% \\
\hline
\hline
\end{tabular}
\vspace{-0.5cm}
\end{center}
\end{table}

\section{Results}
\label{section:Result}

We find no event excess over background predictions in the $p \rightarrow \overline{\nu} K^{+}$ search. The lower limit on the partial proton lifetime is obtained from:
\begin{eqnarray}
\label{equation:lifetime}
\tau / B(p \rightarrow \overline{\nu} K^{+}) & > & \frac{N_{p}}{n_{limit}} \sum_{i = 1, 2} T_{i} \epsilon_{i}
\end{eqnarray}
where $N_{p}$ is the number of protons in the whole LS volume, $T_{i}$ is the detector live time, $\epsilon_{i}$ is the detection efficiency for \mbox{period I} ($i$=$1$) and \mbox{period II} ($i$=$2$) given in Table~\ref{table:efficiency}, and $n_{limit}$ is the estimated upper limit on the number of events for the \mbox{$p \rightarrow \overline{\nu} K^{+}$} signal. Considering the large uncertainty of the background estimate, we conservatively calculate $n_{limit}$ assuming zero background. Using the Feldman-Cousins procedure~\cite{Feldman1998}, we calculate the confidence interval which accounts for the errors on nuisance parameters listed in Table~\ref{table:systematic}, resulting in \mbox{$n_{limit} = 2.44\,{\rm events}$} at 90\% C.L. and the corresponding lifetime limit of \mbox{$\tau / B(p \rightarrow \overline{\nu} K^{+}) > 5.4 \times 10^{32}\,{\rm years}$}.

\section{Discussion}
\label{section:Discussion}

We demonstrated a sensitive search for the proton decay mode $p \rightarrow \overline{\nu} K^{+}$ with a new search method using a large liquid scintillator detector. While the search sensitivity in KamLAND is about an order of magnitude smaller than a previous search in Super-Kamiokande due to the more limited exposure, we confirm the conclusion that the SUSY SU(5) GUT with minimal assumptions predicting a partial lifetime of $\tau / B(p \rightarrow \overline{\nu} K^{+}) \lesssim 10^{30}\,\rm{years}$ is excluded. Extended SUSY-GUT models giving a partial lifetime of $\tau / B(p \rightarrow \overline{\nu} K^{+}) \gtrsim 10^{33}\,\rm{years}$ are still beyond our search range. However, our search sensitivity has not been limited by  backgrounds yet due to the effective background suppression of the coincidence method. It indicates that future increases in statistics are possible with LS detectors.

Several future tens-of-kton LS detectors at different underground sites are proposed to investigate various physics topics. The LENA experiment was planned to perform a sensitive search for $p \rightarrow \overline{\nu} K^{+}$, and the potential of a LS detector with a fiducial mass of 50\,kton was investigated~\cite{Undagoitia2005}. Assuming a projected detection efficiency of 65\%, the search sensitivity is expected to reach \mbox{$\tau / B(p \rightarrow \overline{\nu} K^{+}) > 4 \times 10^{34}\,\rm{years}$} at 90\% C.L. within ten years~\cite{Undagoitia2005}, allowing the extended SUSY-GUT models to be tested. Similarly, the sensitivity for the planned JUNO experiment with a fiducial mass of 20\,kton LS is expected to reach \mbox{$\tau / B(p \rightarrow \overline{\nu} K^{+}) > 1.9 \times 10^{34}\,\rm{years}$} at 90\% C.L. within ten years~\cite{An2015}. Even if a detection efficiency of only 44\% achieved in this study is applied to the searches in LENA and JUNO, the sensitivities are \mbox{$\tau / B(p \rightarrow \overline{\nu} K^{+}) > 2.7 \times 10^{34}\,\rm{years}$} and \mbox{$\tau / B(p \rightarrow \overline{\nu} K^{+}) > 1.5 \times 10^{34}\,\rm{years}$} at 90\% C.L., respectively, which will still significantly impact models. The higher efficiency of 65\% estimated in Ref.~\cite{Undagoitia2005} is mainly due to the clear discrimination of atmospheric neutrino backgrounds based on pulse rise time measurements. From atmospheric neutrino MC studies, we find that the rise time stability in KamLAND is insufficient to provide efficient discrimination. In the cases of LENA and JUNO, kaon production induced by atmospheric neutrino interactions, discussed in Sec.~\ref{subsection:Simulation_atmospheric_neutrino}, are non-negligible. However, if the background discrimination by counting the number of decay electrons suggested in Ref.~\cite{Undagoitia2005} works, a background-free search will still be possible.

\section{Conclusion}
\label{section:Conclusion}

We have presented a search for the proton decay mode $p \rightarrow \overline{\nu} K^{+}$ with the KamLAND liquid scintillator detector. We found no evidence of proton decay and obtained a lower limit on the partial proton lifetime $\tau / B(p \rightarrow \overline{\nu} K^{+}) > 5.4 \times 10^{32}\,\rm{years}$ at 90\% C.L. Due to KamLAND's smaller detector size, the limit is less stringent than the latest Super-Kamiokande result, $\tau / B(p \rightarrow \overline{\nu} K^{+}) > 5.9 \times 10^{33}\,\rm{years}$ at 90\% C.L. However, the detection efficiency of 44\% in KamLAND is about two times better than that of Super-Kamiokande for background-free searches, and the systematic uncertainty of 12\% is almost free from nuclear modeling uncertainties. Based on these results, we expect future searches with tens-of-kton liquid scintillator detectors to be competitive with the water Cherenkov detectors. Such improved searches will test SUSY-GUT models more stringently.

\section*{ACKNOWLEDGMENTS}

We thank T. Kitagaki for advice and guidance. The \mbox{KamLAND} experiment is supported by JSPS KAKENHI Grants No. 16002002 and No. 21000001; 
the World Premier International Research Center Initiative (WPI Initiative), MEXT, Japan; Stichting FOM in the Netherlands; and under the U.S. Department of Energy (DOE) Grants No. DE-AC02-05CH11231 and No. DE-FG02-01ER41166, as well as other DOE grants to individual institutions. The Kamioka Mining and Smelting Company has provided service for activities in the mine. We acknowledge the support of NII for SINET4.

\bibliography{ProtonDecay}

\begin{thebibliography}{10}

\bibitem{Wess1974}
J.~Wess and B.~Zumino,
\newblock Nucl. Phys. B {\bf 70}, 39 (1974).

\bibitem{Marciano1982}
W.~Marciano and G.~Senjanovi{\'c},
\newblock Phys. Rev. D {\bf 25}, 3092 (1982).

\bibitem{Murayama2002}
H.~Murayama and A.~Pierce,
\newblock Phys. Rev. D {\bf 65}, 055009 (2002).

\bibitem{Hirata1989}
K.~S. Hirata {\em et~al.},
\newblock Phys. Lett. B {\bf 220}, 308 (1989).

\bibitem{McGrew1999}
C.~McGrew {\em et~al.},
\newblock Phys. Rev. D {\bf 59}, 052004 (1999).

\bibitem{Nishino2009}
H.~Nishino {\em et~al.}, (Super-Kamiokande Collaboration),
\newblock Phys. Rev. Lett. {\bf 102}, 141801 (2009).

\bibitem{Hayato1999}
Y.~Hayato {\em et~al.}, (Super-Kamiokande Collaboration),
\newblock Phys. Rev. Lett. {\bf 83}, 1529 (1999).

\bibitem{Kobayashi2005}
K.~Kobayashi {\em et~al.}, (Super-Kamiokande Collaboration),
\newblock Phys. Rev. D {\bf 72}, 052007 (2005).

\bibitem{Abe2014b}
K.~Abe {\em et~al.}, (Super-Kamiokande Collaboration),
\newblock Phys. Rev. D {\bf 90}, 072005 (2014).

\bibitem{Svoboda2003}
{R. Svoboda, in Talk at the Eighth International Workshop on Topics in
  Astroparticle and Underground Physics (TAUP), Seattle, WA, 2003
  (unpublished).}

\bibitem{Undagoitia2005}
T.~M. Undagoitia {\em et~al.},
\newblock Phys. Rev. D {\bf 72}, 075014 (2005).

\bibitem{Gando2013b}
A.~Gando {\em et~al.}, (KamLAND Collaboration),
\newblock Phys. Rev. D {\bf 88}, 033001 (2013).

\bibitem{Gando2013a}
A.~Gando {\em et~al.}, (KamLAND-Zen Collaboration),
\newblock Phys. Rev. Lett. {\bf 110}, 062502 (2013).

\bibitem{Andreopoulos2010}
C.~Andreopoulos {\em et~al.},
\newblock Nucl. Instrum. and Methods Phys. Res. Sect. A {\bf 614}, 87 (2010).

\bibitem{Benhar2005}
O.~Benhar, N.~Farina, H.~Nakamura, M.~Sakuda, and R.~Seki,
\newblock Phys. Rev. D {\bf 72}, 053005 (2005).

\bibitem{Hoffmann1995}
M.~Hoffmann, J.~W. Durso, K.~Holinde, B.~C. Pearce, and J.~Speth,
\newblock Nucl. Phys. A {\bf 593}, 341 (1995).

\bibitem{Tarasov2008}
V.~E. Tarasov, V.~V. Khabarov, A.~E. Kudryavtsev, and V.~M. Weinberg,
\newblock Phys. At. Nucl. {\bf 71}, 1410 (2008).

\bibitem{Allison2006}
J.~Allison {\em et~al.},
\newblock IEEE Trans. Nucl. Sci. {\bf 53}, 270 (2006).

\bibitem{Honda2011}
M.~Honda, T.~Kajita, K.~Kasahara, and S.~Midorikawa,
\newblock Phys. Rev. D {\bf 83}, 123001 (2011).

\bibitem{Olive2014}
K. A. Olive {\it et al.} (Particle Data Group), Chin. Phys. C {\bf 38} (2014).

\bibitem{Rein1981}
D.~Rein and L.~M. Sehgal,
\newblock Ann. Phys. (N.Y.) {\bf 133}, 79 (1981).

\bibitem{Alam2010}
M.~Rafi~Alam, I.~Ruiz~Simo, M.~Sajjad~Athar, and M.~J. Vicente~Vacas,
\newblock Phys. Rev. D {\bf 82}, 033001 (2010).

\bibitem{Abe2014a}
K.~Abe {\em et~al.}, (T2K Collaboration),
\newblock Phys. Rev. Lett. {\bf 112}, 061802 (2014).

\bibitem{Feldman1998}
G.~J. Feldman and R.~D. Cousins,
\newblock Phys. Rev. D {\bf 57}, 3873 (1998).

\bibitem{An2015}
F.~An {\em et~al.},
\newblock arXiv:1507.05613v1.

\end{thebibliography}

\end{document}